\documentclass[12pt,preprint]{aastex}

\usepackage{emulateapj5}

\newcommand{\src}{MAXI J1659$-$152 }

\shorttitle{Energy dependent power spectral states and origin of
aperiodic variability}

\shortauthors{Yu et al.}

\begin{document}


\title{Energy dependent power spectral states and origin of aperiodic variability in black hole binaries}


\author{Wenfei Yu and Wenda Zhang}
\affil{Key Laboratqory for Research in Galaxies and Cosmology, Shanghai Astronomical Observatory, Chinese Academy of Sciences, 80 Nandan RD, Shanghai, 200030, China. E-mail: wenfei@shao.ac.cn }






\begin{abstract}
We found the black hole candidate \src showed distinct power spectra, i.e., a power-law noise (PLN) vs. band-limited noise (BLN) plus quasi-periodic oscillations (QPOs), below and above about 2 keV respectively, in the observations with the Swift and the RXTE during the 2010 outburst, indicating a high energy  cut-off of the PLN and a low energy cut-off of the BLN and the QPOs around 2 keV. The emergence of the PLN and the fading of the BLN and the QPOs initially took place from below 2 keV when the source entered the hard intermediate state and finally settled in the soft state three weeks later. The evolution was accompanied by the emergence of the disk spectral component and decreases in the amplitudes of variability in the soft X-ray and the hard X-ray bands. Our results indicate that the PLN is associated with the optically thick disk in both hard and intermediate states, and power spectral state is {\it independent} of the X-ray energy spectral state in a broadband view. We suggest that in the hard and the intermediate state, the BLN and the QPOs emerge from the innermost hot flow subjected to Comptonization, while the PLN originates from the optically thick disk further out. The energy cut-offs of the PLN and the BLN or QPOs then follow the temperature of the seed photons from the inner edge of the optically thick disk, while the high frequency cut-off of the PLN follows the orbital frequency of the inner edge of the optically thick disk as well. 
\end{abstract}


\section{Introduction}

Black hole binaries show well-defined correlated X-ray spectral and variability properties \citep{rm06}. According to the picture formed by the observations in the 2--60 keV with the Rossi X-ray Timing Explorer (RXTE) \citep[for example, ][]{homan01,Belloni2005,Klis06}, distinct variability components are shown in the Fourier power spectra. In the soft state, the power spectrum is well-described by a power-law noise (PLN) component, sometimes with a possible break at around 10 Hz; in the hard state, the power spectrum is dominated by one or more band-limited noise (BLN) components at low frequencies and some times quasi-periodic oscillations (QPOs) at higher frequencies; in the intermediate state usually seen between the hard state and the soft state in time sequence, the power spectrum is composed of BLN components and QPOs \citep{Belloni2002}, and possibly with an additional weak PLN at low frequency. The overall variability amplitude has been found strongly correlated with the fraction of the power-law spectral component \citep{Miyamoto94,mu11}. 

Recent studies show most of the X-ray variability is associated with the power-law spectral component \citep{Klis06}, as shown by the increase of the fractional {\it rms} in relation to the photon energy for those BLN and QPOs \citep[see e.g. ][]{zycki06}. However, improved coverage of the soft X-rays down to below 0.5 keV revealed that the disk spectral component can still contribute a fractional {\it rms} of a few tens percent in the hard state in the low frequency Lorentzian component on the time scales of tens of seconds \citep{wilkinsonuttley}. \cite{Miyamoto94} analyzed GINGA ($>$ 2 keV) observations of the black hole candidate GS 1124$-$683. They concluded that the evolving power spectra in the soft state and in the intermediate state that defined later on were consistent with a decomposition of the power spectrum into a BLN component and a PLN component, which was attributed to the power-law and the disk blackbody spectral components, respectively. But such an additive picture does not hold in the same data when the power-law spectral component is larger than 10\% of the total X-ray flux. 

The soft state power spectrum is usually dominated by a PLN, which was suspected as the signature of disk accretion \citep{Miyamoto94,cui97}. \cite{lyubarskii97} showed that the PLN can be modeled as an inward propagation of disk fluctuations. Since the PLN remains the same shape in a wide range of frequencies and the variations are thought to originate from the power-law spectral component in the soft state, the likely source responsible for the flux variations at all frequencies in this spectral state is the disk covered by the disk corona \citep{churazov01}. On the other hand, the BLN components and the QPOs have been only seen in the hard state and in the intermediate (or very high state) in which the non-thermal spectral component exists \citep[see e.g., ][]{homan01,yu03}. The origin of these components is probably due to the transparency for these high frequency variations in the geometrically thick and optically thin disks, as suggested by \cite{churazov01}. In summary, we lack of direct evidence showing that the power-law noise in the hard state or the intermediate state comes from the optically thick disk and the decoupling of the timing components in the energy spectra.  

\src was discovered in 2010 by Swift \citep{Kann2010,Negoro2010}, which was initially triggered as a Gamma-ray burst (GRB) \citep{Mangano2010}. Its orbital period is around two hours based on its X-ray dips, which makes it the black hole binary with the shortest orbital period known \citep{Kuulkers10,Kuulkers12}. X-ray timing and spectral analysis of the RXTE observations suggest that it is indeed a black hole binary candidate \citep{Kalamkar2011}. It reached 0.2--0.3 crab in the RXTE energy band. Base on the relation between the peak luminosity of LMXB transient outburst and the orbital period \citep{Wu2010}, the distance of \src should not be large.

A series of Swift/XRT and RXTE/PCA monitoring observations of \src were performed during the entire outburst period \citep{Kennea2011,mu11,yamaoka12,Kuulkers12}. The XRT power spectra were clearly energy dependent \citep{Kennea2011}. In this paper, we show during the hard intermediate state the power spectra corresponded to distinct black hole X-ray power spectral states below and above 2 keV, respectively. Furthermore, the emergence of the PLN and the decline of the BLN and the QPO components in the soft X-rays below 2 keV occurred at least several weeks before that was seen in the X-rays above 2 keV during the rising phase of its 2010 outburst. Our results provide strong evidence that the PLN in the hard state and the intermediate state is of disk origin, and black hole power spectral state is dependent on which spectral component we are looking at. 

\section{Observation and Data Analysis}
\label{sec:data}

\subsection{Swift XRT data analysis}
\label{sec:xrtdata}

There were altogether more than 60 Target-of-opportunity $Swift$ observations of \src during its 2010 outburst \citep{Kennea2011,Kuulkers12}. We focused on the first 38 observations taken in the `Windowed Time'(WT) mode with a frame time of 1.766 ms between MJD 55464 and MJD 55492, corresponding to observation ID 0043492800 to 0031843008, which allows a comparison with the quasi-simultaneous RXTE/PCA observations \citep{Kalamkar2011,mu11,yamaoka12}.

Following standard approach to eliminate pile-up effect, we determined the inner radius of source region by evaluating distributions of grade 0 and grade 0-2 events \cite[see Appendix of][also the pile-up thread\footnote{http://www.swift.ac.uk/analysis/xrt/pileup.php}]{P.Romano2006}. The annulus with diameters from 20 to 40 pixels was used to evaluate the fraction unaffected by photon pile-up. We chose the smallest-allowed radius in pixels unaffected by piled-up as the inner radius of source region in order to include as many events as possible. For example, the source region was thus chosen by excluding the innermost 7 pixels for the observation 00434928019.

Our Swift data reduction was performed with HEASOFT 6.11 using the sky coordinate reported in \cite{Kennea2010}. Periodic soft X-ray dips were seen in XRT light-curves due to absorbers in the disk \citep{Kuulkers10,Kennea2011,Kuulkers12}, which brought additional low-frequency variability. We removed events that lies within 1300 seconds of the dip (central) times to avoid the effect even for the longest known dips. The XRT spectral files were binned by $grppha$ to ensure minimum photons of 20 per energy bin. XRT timing analysis was performed with $powspec$ in XRONOS version 5.22. We found the Leahy normalized Fourier power decreases significantly above 50 Hz and deviates from 2 expected for a Poisson noise. So we determined the white noise level by averaging the power between 30 Hz and 50 Hz. 


\subsection{RXTE/PCA data analysis}
We analyzed the RXTE/PCA pointed observations in the same period. In our PCA spectral analysis, we made use of RXTE/PCA standard spectral products to study quasi-simultaneous (within 0.5 days) PCA observations using Xspec 12.7.0. We applied the model $wabs*(diskbb+powerlaw)$ in our spectral fit. To account for inconsistent calibration between the PCA and the XRT in our spectral analysis, we scaled PCA spectra with a constant factor and allowed it to vary. A systematic uncertainty of $1\%$ was applied to the spectral fitting. For the XRT spectra, we ignored energy channel below 0.4 keV and above 10 keV; for the PCA spectra, we ignored energy channel below 3 keV, 4.5-5 keV (Xenon L edge) and above 20 keV. For those Swift observations without quasi-simultaneous RXTE observations within 0.5 days, we fit the XRT spectra alone. We made use PCA Single bit or Event mode data (2--60 keV) and generated average power spectra up to 2048 Hz for each segments with custom routines. The white noise level was estimated by averaging power above 1800 Hz. Throughout our power spectral analysis, we quoted source fractional {\it rms} variability by integrating power between 0.01 and 20 Hz.

\section{Results}
\subsection{Evolution of the X-ray spectra and the variability amplitudes}

Based on the study of the RXTE observations \citep{mu11}, the source stayed  in the hard state ( before MJD 55467, not covered by the RXTE observations), the hard intermediate state (MJD 55467--55481), the soft intermediate state ( MJD 55481--55483), and the soft state (MJD 55484--55490) in a period of nearly four weeks. In Figure 1, we plot the evolution of the spectral and timing properties during the rising phase of its outburst. In our XRT data analysis, we found the fractional {\it rms} variability below 2 keV decreased from 23\% to about 5--8\% around MJD 55467--55468, which was accompanied by an increase of the inner disk temperature to above 0.5 keV and the disk fraction below 2 keV to above 30\%. The PLN started to dominate in the energy band below 2 keV about three weeks earlier than in the entire energy band at around MJD 55490. Our analysis of the UVOT data show that the UVW2 flux rose to its peak right before MJD 55468 and then declined. The rise led the soft X-ray (2--4 keV) rise but lag the hard X-ray (15--50 keV) rise, suggesting the UV emission does not entirely belong to the power-law spectral component. The details of our study of the UVOT data is out of the scope of the paper and will be presented elsewhere. 

\subsection{Distinct power spectral states below and above 2 keV}
\begin{figure}
\epsscale{.80}
 \plotone{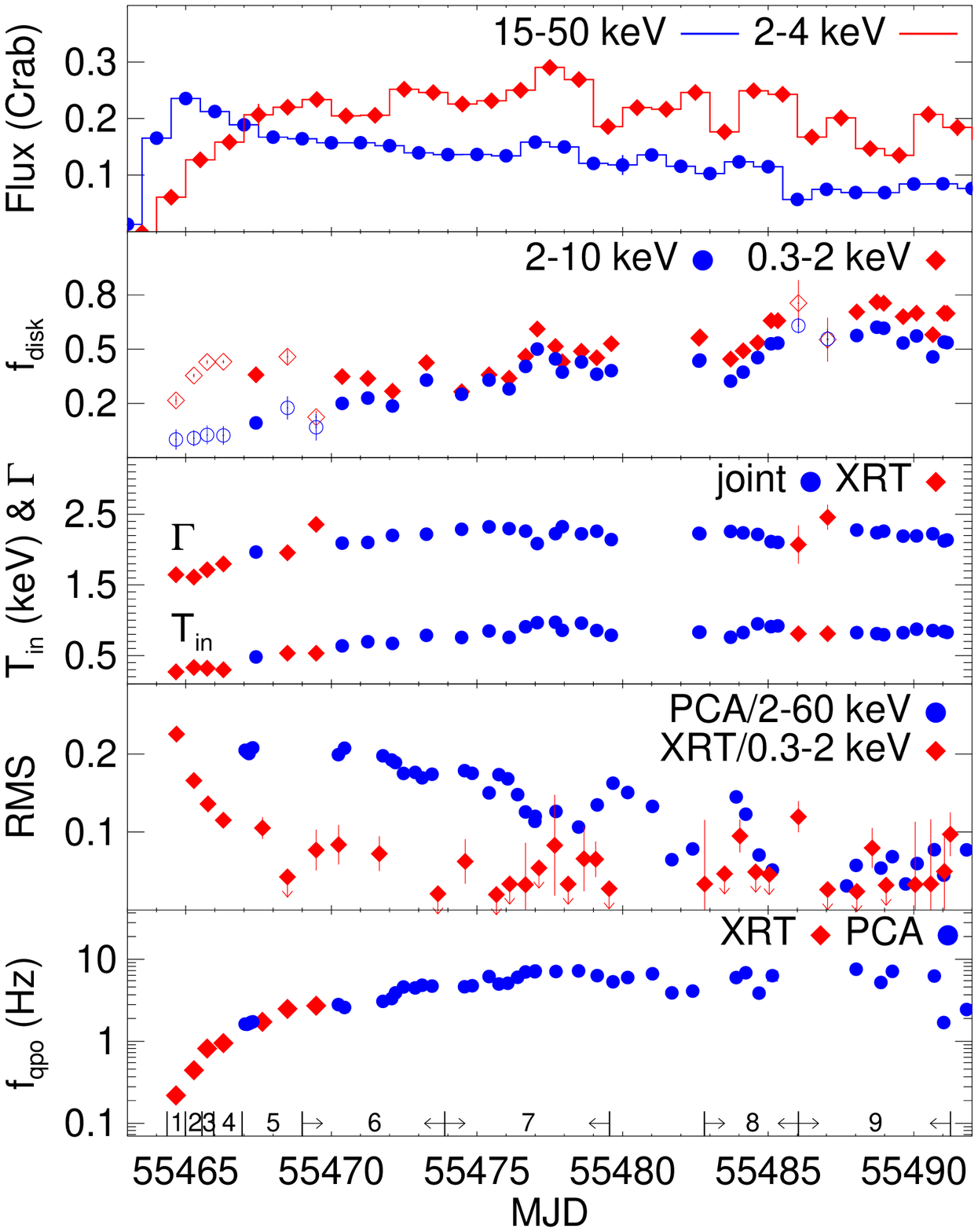} 
 \caption{Evolution of the X-ray intensity as seen with the MAXI (2--4 keV) and the BAT (15--50 keV), the disk energy flux fraction in 0.3--2.0 keV and 2.0--10.0 keV, the inner disk temperature and the power-law photon index measured with the XRT (or joint with the RXTE) observations, the fractional {\it rms} amplitudes in the 0.3--2.0 keV or the 2.0--60 keV band (integrated in 0.01--20 Hz), and the frequency of the primary QPO. Notice that the fractional {\it rms} variability declined much faster in 0.3--2.0 keV than in 2.0--60 keV when the disk black body temperature increased to above 0.5 keV. We found the PLN started to dominate the power spectra in the 0.3--2.0 keV band when the disk spectral component increased to above $\sim$ 30\% (interval 5 and 6) or $\sim$ 50\% in 2.0--60 keV (interval 9). In fact the disk flux fraction in 0.3--2.0 keV should be underestimated since there should be a low energy cut-off of the power-law spectral component to mimic the Comptonization. \label{fig:rms}}
\end{figure}

\begin{figure}
\epsscale{0.6}
\plotone{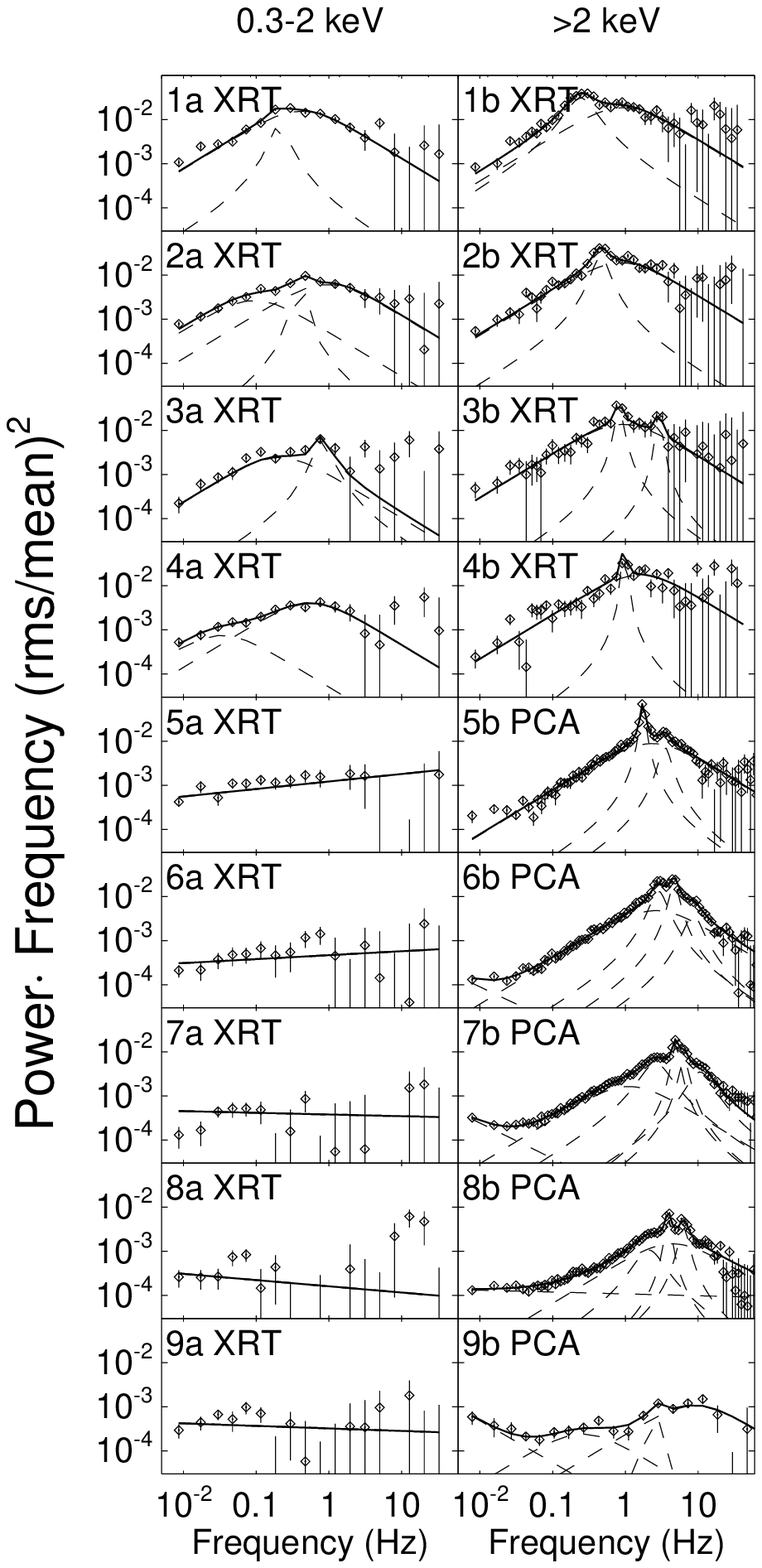}
\caption{Evolution of the average power spectra during the nine time intervals in 0.3--2.0 keV as seen with the Swift/XRT (left column). A PLN dominated in the Interval 5 (see Figure 5a). 
\label{fig:pds1}}
\end{figure}

\begin{figure}
\epsscale{0.6}
\plotone{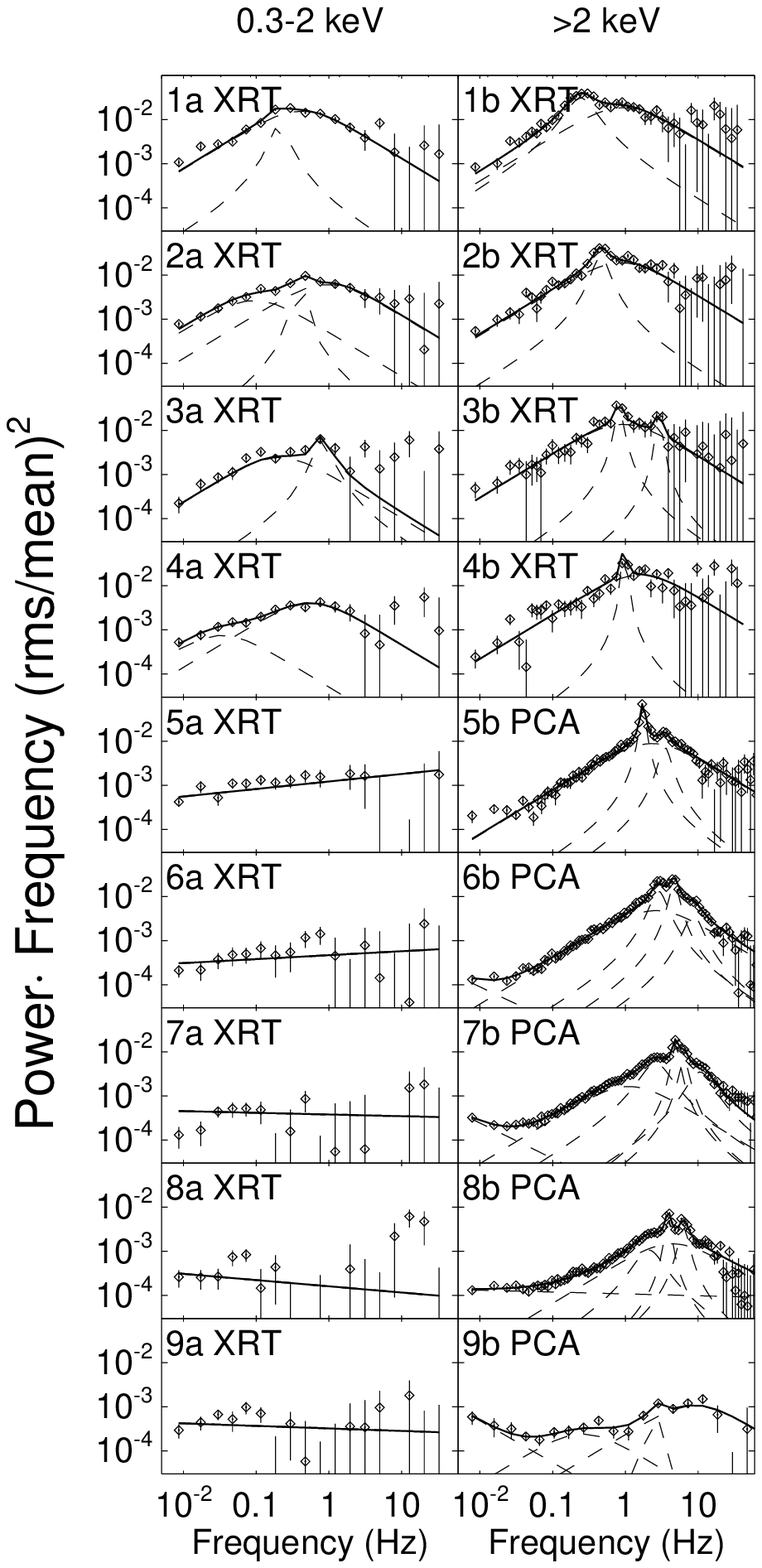}
\caption{Evolution of the average power spectra during the nine time intervals above 2 keV as seen with the RXTE/PCA and the Swift/XRT (right column). Power spectra are either from the XRT (2--10 keV) or from the PCA (2--60 keV). 
\label{fig:pds2}}
\end{figure}

We compared the power spectra in the 0.3--2 keV range seen with the Swift/XRT and in the 2--60 keV range seen with the RXTE/PCA (or in the 2--10 keV range of the XRT when the quasi-simultaneous PCA observations were not available). In order to see the trend clearly, we averaged the corresponding power spectra in nine intervals between MJD 55464--55490, thus determined by similarities of the power spectra. The scheme is shown in the lowest panel of Figure~\ref{fig:rms}. The evolution of the power spectra is shown in Figure~\ref{fig:pds1} and Figure~\ref{fig:pds2}, corresponding to the energy range 0.3--2.0 keV and the 2.0-60.0 keV, respectively. 

All the white-noise subtracted power spectra were fit with models composed of a power-law noise, zero-centered Lorenztians for BLN components, and Lorentzians for QPOs. The uncertainties of the parameters correspond to $\delta \chi^2=1$. In sequence, the power spectra in the energy range 0.3--2.0 keV shows a distinct transition from a characteristic hard state power spectrum dominated by BLN and QPOs (see Figure 2, 4a) to a characteristic soft state power spectra showing only a PLN at around MJD 55467 (See Figure 2, 5a). Notice that the transition occurred when the disk black body temperature increased from about 0.3 keV to above 0.5 keV. 

Two phenomena were seen then. One is the disk fraction exceeded $\sim$30\% in the 0.3--2.0 keV range (actual disk fraction should be higher because the power-law model should have a low energy cut-off to mimic Comptonization). The PLN became insignificant at energies somewhere above 2 keV. In one observation around MJD 55467-55468 (Observation ID 00434928005), the PLN in the 0.3--2.0 keV was about 5 times larger than that seen with the quasi-simultanous RXTE/PCA observation (Observation 95358-01-02-00) in 2--60 keV (Figure ~\ref{fig:plpds}). In order to determine the energy at which the PLN cuts off, we obtained the power spectrum of the XRT data in the 2--4 keV range. We found if we use a model composed of a PLN and several Lorentzians for the BLN and the QPO components, which were fixed at the frequencies of the BLN and QPO components seen in the 2--60 keV PCA data, the best-fit model gives the amplitude of the 2--4 keV PLN (although itself has only a 1 $\sigma$ significance) being 30\% lower than the amplitude of the PLN in the 0.3--2.0 keV. Therefore the PLN component should have a high energy cut-off at around 2--4 keV. How high the energy cut-off of the PLN can be ? We could only put an 1 $\sigma$ upper limit of the PLN in 2.0-4.0 keV, which is weaker than that measured in 0.3-2.0 keV but is still considered as comparable to the 10\% level (0.01-20 Hz range) in the 0.3--2.0 keV band. We found the average XRT photon energy in the 0.3-2.0 keV and the 2.0-4.0 keV XRT bands were 1.3 keV and 2.8 keV, respectively. Both energies are in the energy range of the disk blackbody spectral component with significant contribution to the energy spectrum. The XRT observation therefore put the PLN cut-off energy to roughly above around 2.8 keV. 

The RXTE/PCA observation give an additional constraint on the PLN cut-off energy. We found the 2--60 keV RXTE/PCA observation showed a PLN of about 5 times lower than that seen in the 0.3--2.0 keV with the Swift/XRT. Based on our study of the XRT power spectra in the two energy bands 0.3--2.0 keV and 2.0--4.0 keV, it is likely that only some low energy photons contribute to the PLN (not the entire 2--60 keV photons). We can then estimate the energy of the high energy cut-off of the PLN based on the assumption that only part of the photons contribute to the PLN. We studied the PCA energy spectrum and found 20\% PCA photons were below 3.5 keV, 40\% PCA photons were below 4.7 keV, and 50\% photons had energies less than 7.4 keV. Therefore, assuming the PLN energy spectrum is a step function (between 10\% rms at low energies as in 0.3-2.0 keV and 0\% rms above a cut-off energy), if only the photons below 3.5 keV contribute to the PLN at the same amplitude level as seen in the 0.3-2.0 keV with the XRT, the overall amplitude of the PLN in the 2--60 keV PCA band would be 5 times lower than that seen in 0.3--2.0 keV, as what we observed with the PCA. This will put a constraint on the PLN cut-off energy at most at 3.5 keV (notice that the PCA energy resolution is poor so the cut-off energy should be approximate); if the soft photons contribute to the PLN at half of the amplitude as seen in the 0.3--2.0 keV XRT band, the highest energy photon contributes to the PLN would reach 4.7 keV but the cut-off energy should be lower than 3.5 keV (in this case 40\% PCA photons with the lowest energies contribute to the PLN). This is a very conservative estimate since the PLN in the 2-4 keV as seen with the XRT probably has an amplitude probably comparable to that seen in the 0.3--2.0 keV. In the XRT energy spectrum, we found at 4.7 keV the disk blackbody component in a disk blackbody plus a power-law model contributed 10\% of the total photon counts, while at around 1.3 keV the disk blackbody component contributed 30\% of the total photon counts. So the PLN energy spectrum does look more like a disk blackbody than a steep power-law, with a high energy cut-off that is consistent with the disk spectral component. 

The other phenomenon associated with the transition between Interval 4 and 5 is that the BLN and QPOs disappeared in the soft band 0.3--2.0 keV, indicating a low energy cut-off of the BLN and the QPOs, while the power spectra in the energy range above 2.0 keV showed characteristic BLN and QPOs in the hard or the intermediate state up to MJD 55486 (Figure 3, 4b--8b). We know in interval 8 the source transited to the soft state for a while \citep{mu11}. Therefore in the X-rays above 2 keV, the transition of power spectrum to a PLN took place nearly 3 weeks after that occurred in 0.3--2.0 keV.

We have also investigated the relative strength of the BLN and the PLN in the soft X-ray band 0.3--2.0 keV. For the power spectrum in the soft band in each intervals 5--9, we fixed the PLN slope to an power-law index of -1 and the width of the zero-centered Lorentzian corresponding to the width of the BLN component obtained with the simultaneous 2--60 keV RXTE power spectrum. The normalizations of the PLN and the BLN were relaxed. We found there was no need to include a BLN component in the fits, and therefore we can only put upper limits on potential BLN variability amplitudes in the 0.3--2.0 keV range. In the end, we found the amplitude of the PLN was always more than two times of the BLN upper limit, which was in the range between 1.1\% to 3.6\%. The amplitude of the BLN components then rose sharply above about 2 keV to more than 10\% (interval 5--7). This demonstrates that distinct power spectral states were seen in 0.3--2.0 keV and above 2.0 keV (2--60 keV PCA band), as shown in 5a and 5b or in 6a and 6b of Figure~\ref{fig:pds1} and Figure~\ref{fig:pds2}, indicating a transition of the power spectral state across the soft and the hard bands around MJD 55470 -- the energy boundary which divides the X-ray spectrum into disk dominated and power-law dominated regimes. In Figure ~\ref{fig:mulpds}, we plot the average power spectra corresponding to the intervals 5 and 6. It hints a frequency cut-off of the PLN seen in the 0.3--2.0 keV energy range at around 1 Hz, overlapping with the frequencies of the BLN and the QPOs seen at above 2 keV in the PCA band.  We found the deficiency of the PLN power above 1 Hz has nothing to do with the subtraction of the white-noise level. Both energy cut-offs ($\sim$ 2--3.5  keV) of the PLN and BLN components and the possible frequency cut-off ($\sim$ 1 Hz) of the PLN imply that the PLN and the BLN plus QPOs were almost  separated in energy and in frequency during those observations. 

\section{ Conclusion \& Discussion}

The relatively slow evolution of \src between the spectral and timing states provided us an opportunity to zoom in how the spectral content of the timing components evolves in black hole transients. With the help of both Swift/XRT and RXTE/PCA, we are able to look at the energy-dependence of the power spectra in much more details. Although an energy-frequency 2D spectral space was not mapped out by the observations, these observations provided an important picture of the energy dependence of the properties of black hole power spectral states. 

In the conventional picture of black hole spectral states, there is a tight correlation between the energy spectra and the power spectra shape \citep{homan01,rm06}. This is incorrect when we extend the power spectral study down to soft energies. We found \src showed a PLN in the soft X-ray energy band dominated by the thermal spectral component when it just transited from the hard state to the intermediate state, while at the same time displaying typical BLN and QPOs in the hard X-ray energy band dominated by the power-law spectral component. Therefore we found distinct power spectral states, i.e., a PLN vs. BLN plus QPOs, coexisted below and above a certain cut-off energy estimated in the range 2--3.5 keV, suggesting that the conventional black hole state picture we have is primarily a hard X-ray view based on the RXTE/PCA band, and a broadband (or even) a multi-wavelength view is needed to better understand black hole power-spectral states. 


The simultaneous emergence of the PLN and the thermal disk spectral component from below 2 keV provides strong evidence that in the hard or the intermediate state, the PLN is from the approaching disk flow. The disappearance of the BLN and QPOs and the decrease of the overall variability amplitude started in the soft X-ray band below 2 keV as well, which indicates that the photons responsible for the BLN and the QPOs emerged  from regions other than the optically thick disk. A reasonable picture is that the photons responsible for the BLN and the QPOs originated from the innermost disk flow, which were up-scattered in the corona. This could naturally contribute to some hard lag at low frequencies in the black hole hard state, but may be not the major cause of the lag as have been suggested to come from propagation of the low frequency variability into the hot coronal flow \citep{uttley}. The separation of the simultaneous PLN and BLN (or the QPOs) in energy and in frequency might be set up by the inner disk edge disrupted by the hot coronal flow. Comptonization of the soft seed photons originated from the inner edge of the optically thick disk. Because they were up-scattered to higher energies, these photons did not contribute to the PLN seen in the soft X-rays,  but on the other hand, contribute to the BLNs and QPOs, leading to a high frequency cut-off of the PLN at the BLN or QPO frequencies. We might have seen such a cut-off at around 1 Hz where the BLN and the QPOs were peaked (see Figure 6). The PLN in the intermediate state should evolve into the PLN seen in the soft state, which is probably accompanied by the formation of a disk corona that is needed to explain the energy spectrum in the soft state \citep{churazov01}. 

In summary, \src has provided the best example of an isolation rather than a coupling among the power spectral components in the energy spectrum, which revealed clues to the origin of the aperiodic variability in black hole binaries: the observations provided the evidence that the PLN is of disk origin and the BLN and QPOs are of coronal origin, which have been realized for quite some years but has been lack of direct evidence.  Similar result on the BLN component but not those of the emergence of the PLN has recently been seen in SWIFT J1753.5-0127 \citep{kalamkar13}, which supports our conclusion. Our work also further suggests that there might be links in the characteristic frequency and in the (seed) photons between the PLN and the BLN (or QPOs). This is a very important problem to explore in the future since our current power spectral modeling is based on superposition of noise components and QPOs of some mathematical forms (power-law noise and Lorentzians). Justifications of the mathematical forms for the noise and QPO components as well as the simple superposition method are very necessary towards a physical modeling of the black hole power spectrum. 

\acknowledgments

We would like to thank the {\it Swift} operation team for scheduling the TOO observations and a large international team for requesting these TOO observations that we analyzed in this paper. We would also like to thank the anonymous referee for useful suggestions which help improve this work significantly. WY acknowledge useful discussions with Michiel van der Klis, Phillip Kaaret, William Zhang, Mike Nowak, Tod Strohmayer, and Deepto Chakrabarty. We acknowledge the use of data obtained through the High Energy Astrophysics Science Archive Research Center Online Service, provided by the NASA/Goddard Space Flight Center. This work is supported by the National Basic Research Program of China (973 project 2009CB824800) and the National Natural Science Foundation of China (10833002,11073043).

\clearpage

\begin{figure}
\epsscale{.80} 
\plotone{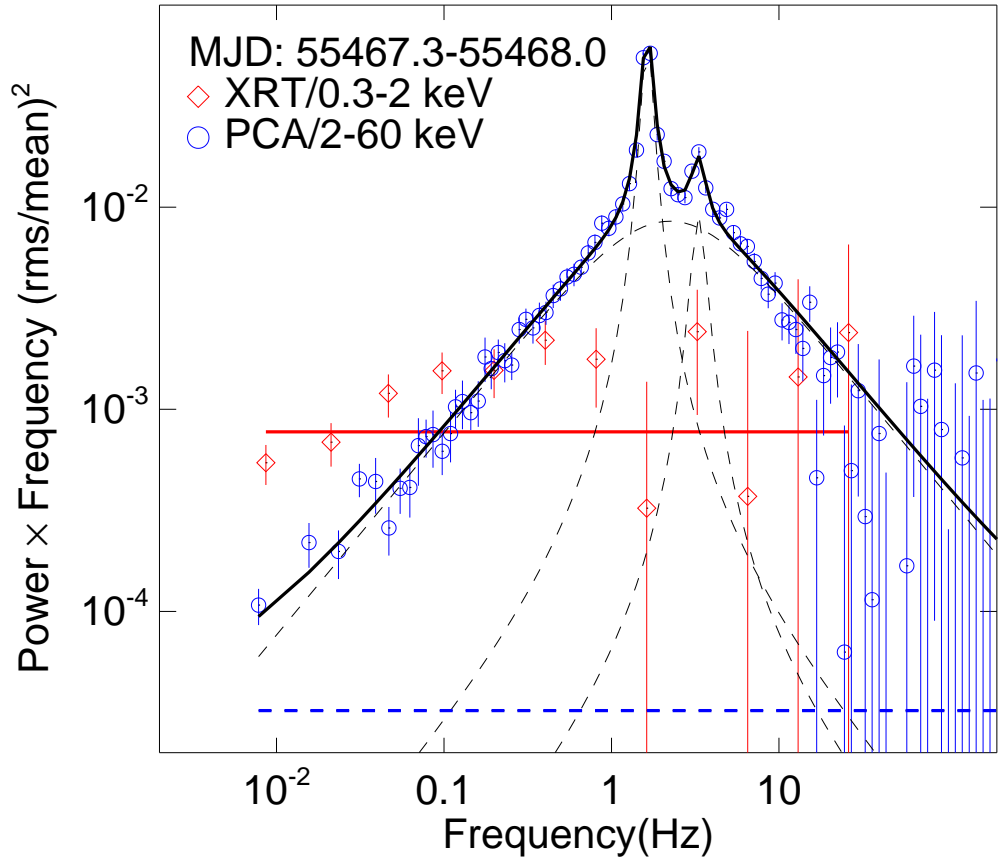} 
\caption{Distinct power-law noise (PLN) below and above 2 keV during single observations during MJD 55467--55468 (Swift observation 00434928005 and RXTE observation 95358-01-02-00). The {\it  rms} of the PLN above 2 keV measured with the RXTE/PCA was 4.9 times lower than that of the PLN in the 0.3--2.0 keV ( its 3 $\sigma$ {\it rms} upper limit was still about 3 times lower than that of the PLN seen in the 0.3--2.0 keV). The red line and the blue dashed-line indicate the best-fit PLN models for the 0.3--2.0 keV and the 2--60 keV data with a power-law index of -1. This indicates a high energy cut-off of the PLN and a low-energy cut-off of the BLN and QPOs.  
\label{fig:plpds}}
\end{figure}

\begin{figure}
\epsscale{.80} 
\plotone{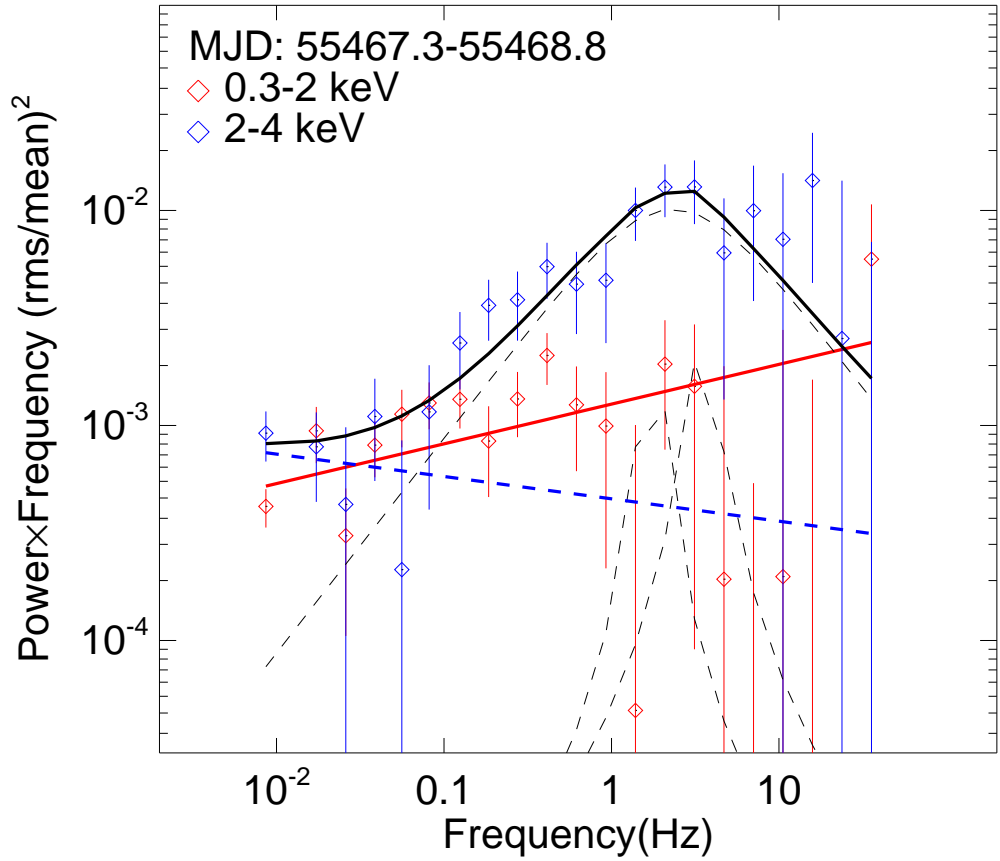} 
\caption{The power spectra in 0.3--2.0 keV (red) and 2.0--4.0 keV (blue) averaged over two Swift observations during MJD 55467--55468 (Interval 5). The {\it  rms} of the PLN in the 0.3--2.0 keV was 9.4$\pm$ 0.9 \%, while in 2.0--4.0 keV the 1-$\sigma$ {\it rms} upper limit of a PLN with a power-law index of -1 was 8.4\%. The best-fit PLN corresponding to the 0.3--2.0 keV power spectrum is shown as a red line. The best-fit PLN corresponding to the 2.0--4.0 keV power spectrum  is shown as a blue dashed line ( the BLNs (or QPOs) were fixed to the frequencies determined in the 2--60 keV with RXTE/PCA data). This hints a high energy cut-off of the PLN and a low energy cut-off of the BLN and the QPOs. 
\label{fig:plpds}}
\end{figure}

\begin{figure}
\epsscale{.80} 
\plotone{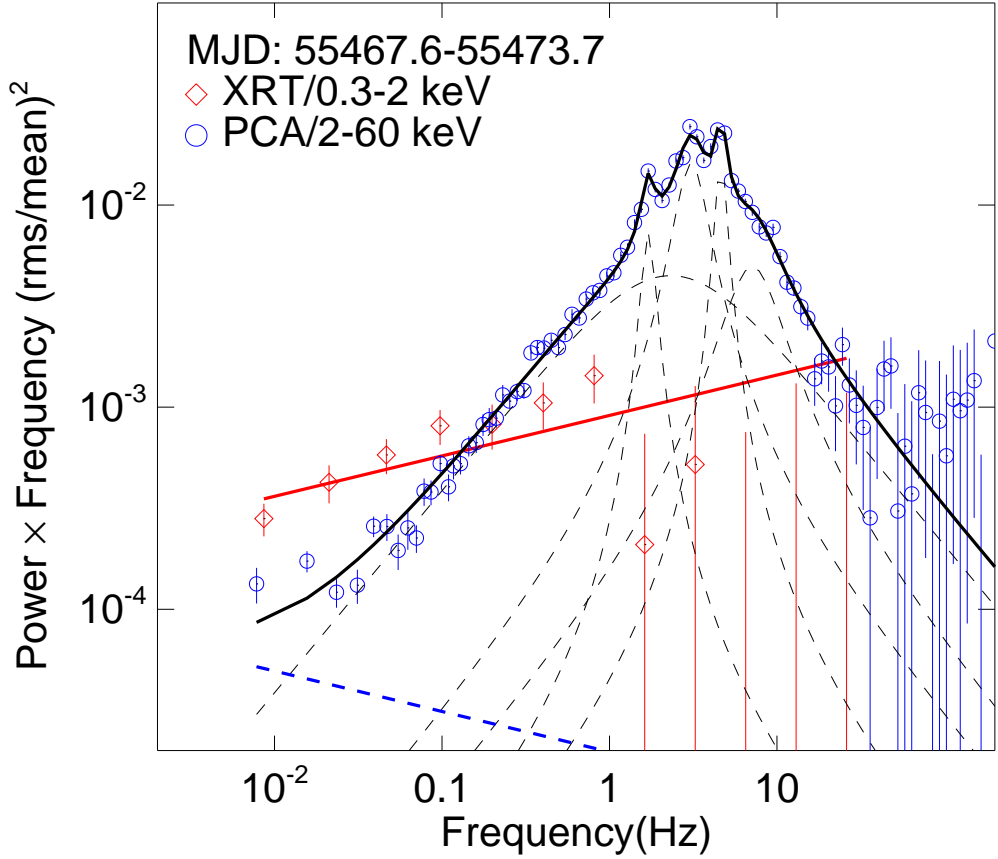} 
\caption{Distinct average power spectra below and above 2 keV seen simultaneously during MJD 55467--55474 (intervals 5 and 6). The power spectrum below 2 keV obtained with the Swift/XRT (in red) shows a single power-law noise (PLN) component with a likely cut-off at about 1 Hz, but the corresponding power spectrum above 2 keV obtained with the RXTE/PCA (in blue) mainly consists of band-limited noise (BLN) and QPOs above 1 Hz. Solid lines represent the best fit models, in which the power-law index of the PLN was allowed to vary to a range between -0.8 and -1.2. The dashed lines represents BLN or QPO components as well as the power-law component for the average power spectrum above 2 keV obtained with the RXTE/PCA. Notice the QPO frequency evolved dramatically in the 6-day interval, which led to multiple peaks in the average power spectrum in the hard X-ray band. The best-fit PLN component in 2--60 keV (blue dashed line) has a much lower amplitude than that in the 0.3--2.0 keV range (red line). The simultaneous power spectra indicate a strong energy dependence of the PLN and the BLN or QPO components.  
\label{fig:mulpds}}
\end{figure}



\begin{thebibliography}{}

\bibitem[Belloni et al.(2005)]{Belloni2005}
Belloni, T., Homan, J., Casella, P., van der Klis, M., Nespoli, E.,
  Lewin, W.~H.~G., Miller, J.~M., \& M{\'e}ndez, M. 2005, \aap, 440, 207

\bibitem[Belloni et al.(2002)]{Belloni2002}
Belloni, T., Psaltis, D., \& van der Klis, M. 2002, \apj, 572, 392

\bibitem[Belloni (2010)]{Belloni2010}
Belloni, T. M. 2010, in Lecture Notes in Physics, Berlin Springer Verlag,  Vol. 794, Lecture Notes in Physics, Berlin Springer Verlag,
ed. T. Belloni, 53

\bibitem[Churazov et al.(2001)]{churazov01} Churazov, E., Gilfanov, M., \& Revnivtsev, M.\ 2001, \mnras, 321, 759

\bibitem[Cui et al.(1997)]{cui97} Cui, W., Heindl, W. A.,
Rothschild, R. E., et al.\ 1997, \apjl, 474, L57

\bibitem[Gandhi(2009)]{gandhi09} Gandhi, P.\ 2009, \apjl, 697,
L167

\bibitem[Homan et al.(2001)]{homan01} Homan, J., Wijnands, R.,
van der Klis, M., et al.\ 2001, \apjs, 132, 377

\bibitem[Kalamkar et al.(2011)]{Kalamkar2011}
Kalamkar, M., Homan, J., Altamirano, D., van~der Klis, M., Casella,
P., \&  Linares, M. 2011, The Astrophysical Journal Letters, 731, L2

\bibitem[Kalamkar et al.(2013)]{kalamkar13} Kalamkar, M., van der Klis, M., Uttley, P., Altamirano, D., \& Wijnands, R.\ 2013, \apj, 766, 89 

\bibitem[{{Kann}(2010)}]{Kann2010}
Kann, D. A. 2010, GRB Coordinates Network, 11299, 1

\bibitem[Kennea et al.(2010)]{Kennea2010}
Kennea, J. A., Krimm, H., Mangano, V., Curran, P., Romano, P.,
  Evans, P., \& Burrows, D. N. 2010, The Astronomer's Telegram, 2877, 1

\bibitem[Kennea et al.(2011)]{Kennea2011}
Kennea, J. A., et al. 2011, The Astrophysical Journal, 736, 22

\bibitem[Kuulkers et al.(2010)]{Kuulkers10}
Kuulkers, E., et al. 2010, The Astronomer's Telegram, 2912, 1

\bibitem[Kuulkers et al.(2013)]{Kuulkers12} Kuulkers, E., Kouveliotou, C., Belloni, T., et al.\ 2013, \aap, 552, A32 

\bibitem[Lyubarskii(1997)]{lyubarskii97} Lyubarskii, Y. E.\ 1997, \mnras, 292, 679

\bibitem[Leahy et al.(1983)]{Leahy1983}
Leahy, D., Darbro, W., Elsner, R., Weisskopf, M., Kahn, S.,
  Sutherland, P., \& Grindlay, J. 1983, \apj, 266, 160

\bibitem[Mangano et al.(2010)]{Mangano2010}
Mangano, V., Hoversten, E. A., Markwardt, C. B., Sbarufatti, B.,
  Starling, R. L. C., \& Ukwatta, T.~N. 2010, GRB Coordinates Network,
  11296, 1

\bibitem[Motch et al.(1983)]{Motch83} Motch, C., Ricketts, M. J., Page, C. G., et al. 1983, \aa, 119, 171

Mangano, V., Hoversten, E. A., Markwardt, C. B., Sbarufatti, B.,
  Starling, R. L. C., \& Ukwatta, T.~N. 2010, GRB Coordinates Network,
  11296, 1


\bibitem[McHardy et al.(2006)]{mchardy} McHardy, I. M.,
Koerding, E., Knigge, C., Uttley, P., \& Fender, R. P.\ 2006, \nat,
444, 730

\bibitem[Miyamoto et al.(1992)]{Miyamoto1992}
Miyamoto, S., Kitamoto, S., Iga, S., Negoro, H., \& Terada, K. 1992,
  \apjl, 391, L21

\bibitem[Miyamoto et al.(1994)]{Miyamoto94} Miyamoto, S.,
Kitamoto, S., Iga, S., Hayashida, K., \& Terada, K.\ 1994, \apj,
435, 398

\bibitem[Mu{\~n}oz-Darias et al.(2011)]{mu11} Mu{\~n}oz-Darias, T., Motta, S., Stiele, H., \& Belloni, T. M.\ 2011, \mnras, 415, 292


\bibitem[Negoro et al.(2010)]{Negoro2010}
Negoro, H., et al. 2010, The Astronomer's Telegram, 2873, 1

\bibitem[Psaltis et al.(1999)]{psaltis99} Psaltis, D., Belloni, T., \& van der Klis, M.\ 1999, \apj, 520, 262

\bibitem[Remillard \& McClintock (2006)]{rm06} Remillard, R. A., \& McClintock, J.~E.\ 2006, \araa, 44, 49

\bibitem[P. Romano et al. (2006)]{P.Romano2006}
P. Romano et al. 2006, A\&A, 456, 917

\bibitem[{Remillard \& McClintock (2006)}]{Remillard2006}
Remillard, R. A., \& McClintock, J. E. 2006, Annual Review of
Astronomy and  Astrophysics, 44, 49

\bibitem[Sobolewska \& Zycki (2006)]{zycki06} Sobolewska, M.A. \& Zycki, P.T. 2006 \mnras, 370, 405

\bibitem[Spruit \& Kanbach(2002)]{henk02} Spruit, H. C., \& Kanbach, G.\ 2002, \aap, 391, 225

\bibitem[Uttley et al.(2011)]{uttley} Uttley, P., Wilkinson, T., Cassatella, P., et al. \ 2011, \mnras, 414, L60

\bibitem[{{van der Klis}(2006)}]{Klis06}
van der Klis, M. 2006, Rapid X-ray Variability, ed. Lewin, W. H. G.
\& van der Klis, M., 39--112


\bibitem[Wijnands \& van der Klis(1999)]{wijnands99} Wijnands, R., \& van der Klis, M.\ 1999, \apj, 514, 939

\bibitem[Wilkinson \& Uttley (2009)]{wilkinsonuttley} Wilkinson, T. \& Uttley, P., \ 2009, \mnras, 397, 666

\bibitem[Wu et al.(2010)]{Wu2010}
Wu, Y. X., Yu, W., Li, T. P., Maccarone, T. J., \& Li, X. D. 2010,
The Astrophysical Journal, 718, 620

\bibitem[Yamaoka et al.(2012)]{yamaoka12} Yamaoka, K., Allured, R., Kaaret, P., et al.\ 2012, \pasj, 64, 32


\bibitem[Yu et al.(2003)]{yu03} Yu, W., Klein-Wolt, M.,
Fender, R., \& van der Klis, M.\ 2003, \apjl, 589, L33


\end{thebibliography}
\end{document}